\documentclass[9pt,conference, compsocconf]{IEEEtran}
\pdfoutput=1
\usepackage{booktabs} 
\usepackage[pdftex]{graphicx}
\usepackage{amsmath}
\usepackage{enumitem}
\usepackage{setspace}
\usepackage{array}
\usepackage{etoolbox}
\makeatletter
\patchcmd{\@makecaption}
{\scshape}
{}
{}
{}
\makeatletter
\patchcmd{\@makecaption}
{\\}
{.\ }
{}
{}
\makeatother

\usepackage[margin=0.50in]{geometry}

\usepackage{cite}

\graphicspath{{./}}

\usepackage{url}

\makeatletter
\IEEEtriggercmd{\reset@font\normalfont\fontsize{6.8pt}{8pt}\selectfont}
\makeatother
\IEEEtriggeratref{1}

\makeatletter
\let\old@ps@headings\ps@headings
\let\old@ps@IEEEtitlepagestyle\ps@IEEEtitlepagestyle
\def\confheader#1{%
	\def\ps@headings{%
		\old@ps@headings%
		\def\@oddhead{\strut\hfill#1\hfill\strut}%
		\def\@evenhead{\strut\hfill#1\hfill\strut}%
	}%
	\def\ps@IEEEtitlepagestyle{%
		\old@ps@IEEEtitlepagestyle%
		\def\@oddhead{\strut\hfill#1\hfill\strut}%
		\def\@evenhead{\strut\hfill#1\hfill\strut}%
	}%
	\ps@headings%
}
\makeatother

\confheader{%
	26$^{th}$ IEEE Asian Test Symposium 2017, 
	27-30 Nov 2017, Taipei City, Taiwan
	\qquad DOI: TBD, \copyright~2017 IEEE
}

\begin{document}
	\title{Bringing Fault-Tolerant GigaHertz-Computing to Space\\\vspace{-3pt}{\LARGE A Multi-Stage Software-Side Fault-Tolerance Approach for Miniaturized Spacecraft}\vspace{-6pt}}
	
	\author{\IEEEauthorblockN{Christian M. Fuchs\IEEEauthorrefmark{1},
	Todor P. Stefanov\IEEEauthorrefmark{1},
	Nadia M. Murillo\IEEEauthorrefmark{2} and
	Aske Plaat\IEEEauthorrefmark{1}}
	\IEEEauthorblockA{\IEEEauthorrefmark{1}Leiden Institute for Advanced Computer Science
	\IEEEauthorrefmark{2}Leiden Observatory\\Leiden University, 2333 CA, The Netherlands\\
	Email: c.fuchs@liacs.leidenuniv.nl}
	\vspace{-27pt}}
	

	\maketitle
	
	\IEEEpubid{0000--0000/00\$00.00 \copyright\ 2013 IEEE}
	
	\begin{abstract}
		Modern embedded technology is a driving factor in satellite miniaturization, contributing to a massive boom in satellite launches and a rapidly evolving new space industry.
		Miniaturized satellites, however, suffer from low reliability, as traditional hardware-based fault-tolerance (FT) concepts are ineffective for on-board computers (OBCs) utilizing modern systems-on-a-chip (SoC).
		Therefore, larger satellites continue to rely on proven processors with large feature sizes.
		Software-based concepts have largely been ignored by the space industry as they were researched only in theory, and have not yet reached the level of maturity necessary for implementation.
		We present the first integral, real-world solution to enable fault-tolerant general-purpose computing with modern multiprocessor-SoCs (MPSoCs) for spaceflight, thereby enabling their use in future high-priority space missions.
		The presented multi-stage approach consists of three FT stages, combining coarse-grained thread-level distributed self-validation, FPGA reconfiguration, and mixed criticality to assure long-term FT and excellent scalability for both resource constrained and critical high-priority space missions.
		Early benchmark results indicate a drastic performance increase over state-of-the-art radiation-hard OBC designs and considerably lower software- and hardware development costs.
		This approach was developed for a 4-year European Space Agency (ESA) project, and we are implementing a tiled MPSoC prototype jointly with two industrial partners.
	\end{abstract}
	
	\vspace{-9pt}
	\section{Introduction}
	Modern embedded technology is a driving factor in satellite miniaturization, contributing to a massive boom in satellite launches and a rapidly evolving new space industry.
	Micro- and nanosatellites (100-1kg) have become increasingly popular platforms for a variety of commercial and scientific applications, due to an excellent balance of performance and cost.
	However, this class of spacecraft suffers from low reliability, discouraging its use in long, complex, or high-priority missions.
	The on-board computer (OBC) related electronics constitute a much larger share of a miniaturized satellite than they do in larger satellites.
	Thus, per component, they must deliver drastically better performance and consume less energy.
	Therefore, and due to cost considerations, miniaturized satellite OBCs are generally based upon processors with considerably finer feature size, such as those developed for mobile embedded devices.
	
	Traditional hardware-based fault-tolerance (FT) concepts for general-purpose computing, however, are ineffective for modern, highly scaled systems-on-chip (SoCs), becoming a prime source of malfunctions aboard miniaturized satellites \cite{langer2016reliability}.
	Larger satellites, too, are limited by the measures traditionally used to assure FT for space applications, as these prevent larger satellites from harnessing the benefits of modern processors designs, and multiprocessor-SoCs (MPSoCs).
	Also, these hardware-based FT-measures can not handle varying performance requirements during multi-phased missions and mega-constellations \cite{bastida2016mega}.
	Software-based FT measures rapidly evolved due to efforts of the scientific community, and are effective for modern embedded hardware.	
	However, these advances have largely been ignored by the space industry as they were researched only in theory, but rarely designed to be implemented.
	While many of these concepts include innovative ideas, major implementation obstacles and fundamental issues remain unaddressed.
	Often, prior research makes impractical assumptions towards the platform or application environment, ignores fault detection, recovery from fail-over, or other real-world constraints.
	Many concepts also attempt to uphold safety and availability, e.g., for atmospheric aerospace use, but not computational correctness.
	To the best of our knowledge, no integral and practical solution to utilizing modern MPSoC-based systems within high-priority space missions has been developed to date.

	There is a wide gap between academic research towards novel FT concepts and their practical application in spacecraft OBCs.
	Satellite computers for control purposes are still largely based upon architectures developed decades ago, while theoretical research has not achieved the level of maturity necessary to bridge this gap.
	Thus, neither traditional hardware- nor software-based FT solutions could offer all the functionality necessary to improve the reliability of state-of-the-art embedded SoCs in miniaturized satellite OBCs.
	Other concepts promise excellent FT guarantees in theory, but require complex architectures that often do not address the specific challenges of computers flying in space.
	Innovations are especially needed in general-purpose computing, as OBCs must execute a broad variety of applications efficiently.
	The presented research addresses these challenges and our main contributions are:
	\begin{itemize}[leftmargin=.35cm]
		\item the first non-intrusive, integral, flexible, software-side approach enabling the use of modern MPSoCs for spaceflight meeting real-world constraints;
		
		\item an approach not based upon custom or proprietary FT-processor cores, does not require radiation-hard ASICs, or non-standard functionality;
		
		\item which can be implemented with standard toolchains, commercial off the shelf (COTS) components, library IP, and little manpower;
		
		\item an introduction to an FPGA-based MPSoC architecture developed as an ideal platform for our approach.
	\end{itemize}
	
	This approach was developed for a 4-year European Space Agency (ESA) project with two industrial partners.
	Due to the interdisciplinary nature of this project, other aspects of this approach and its hardware implementation will be presented in separate publications.
	
	In the next two sections, we will outline the challenges faced in the space environment, and related work.
	Section \ref{sec:overview} contains a brief overview of the multi-stage approach, its limitations, terminology, as well as the application model and requirements.
	Each stage is described in the subsequent sections, with the supervision concept explained in Section \ref{sec:supervision}.
	Section \ref{sec:platformarch} then introduces briefly an MPSoC architecture specifically designed as a platform for this FT concept.
	Performance and checkpoint reliability are discussed in Section \ref{sec:discussion}, followed by conclusions.

	\vspace{-9pt}
	\section{The Space Environment}
	\label{sec:space}
	Solar cells are the main power source aboard modern spacecraft.
	A spacecraft's orbit, location and orientation (attitude) relative to the Sun, and the solar array's temperature all influence the efficiency of its solar array.
	Miniaturized satellite's have comparably small solar arrays with strongly fluctuating output, and their OBCs are limited to a few Watts of power-budget.	
	Mid-mission physical access to spacecraft is impossible, and historically servicing missions were conducted only on rare occasions for satellites of outstanding importance in low-Earth orbit (LEO).
	Signal-travel times, limited communication windows, and scarce bandwidth make live-interaction with a spacecraft impractical.
	Thus, faults detected during a satellite mission must be resolved unattended, remotely, and fully autonomously.
	The drastically different fault-model, mass, size, physical stress, and thermal design constraints in space \cite{marinella2013total} prevent the re-use of FT-, debugging-, and testing approaches developed for ground application.
	
	
	High-energy particles are the predominant cause for faults within OBCs \cite{Bourdarie2008}.
	They travel along the Earth's magnetic field-lines in the Van Allen belts, are ejected by the Sun during Solar Particle Events, or arrive as Cosmic Rays from beyond our solar system.
	In LEO, the residual atmosphere and Earth's magnetic field provide some protection from radiation, but this absorption effect diminishes quickly with altitude.
	Hence, microelectronics are exposed to a mix of highly charged particles, with flux density depending on solar activity and the spacecraft's attitude.
	They can corrupt logical operations, induce bit-flips within data-storage cells and connecting circuitry, or cause displacement damage (DD).
	
	The impact of radiation on different microfabrication processes, substrates, and memory technologies varies.
	In general, electronics with a large feature size are more resilient to radiation-induced single event effects (SEEs) than those manufactured in finer production nodes.
	Highly scaled chips are susceptible to multi-bit upsets, propagating within circuits corrupting groups of circuits or memory cells.
	The increased impact of radiation effects on finer feature size chips also prevents better protection through more circuit-level protection.
	
	Radiation events can also cause single event functional interrupts (SEFIs), affecting sets of circuits, individual interfaces, or even entire chips.
	In general, the effects of SEEs and SEFIs can be transient or permanent and accumulative, while DD is always permanent \cite{Schwank2013}.
	The accumulative nature of permanent faults implies accelerated and often spontaneous ageing, which must be handled efficiently throughout the entire mission, sometimes as long as 10+ years.
	
\begin{figure}[!t]
		\centering
		\includegraphics[width=0.75\linewidth]{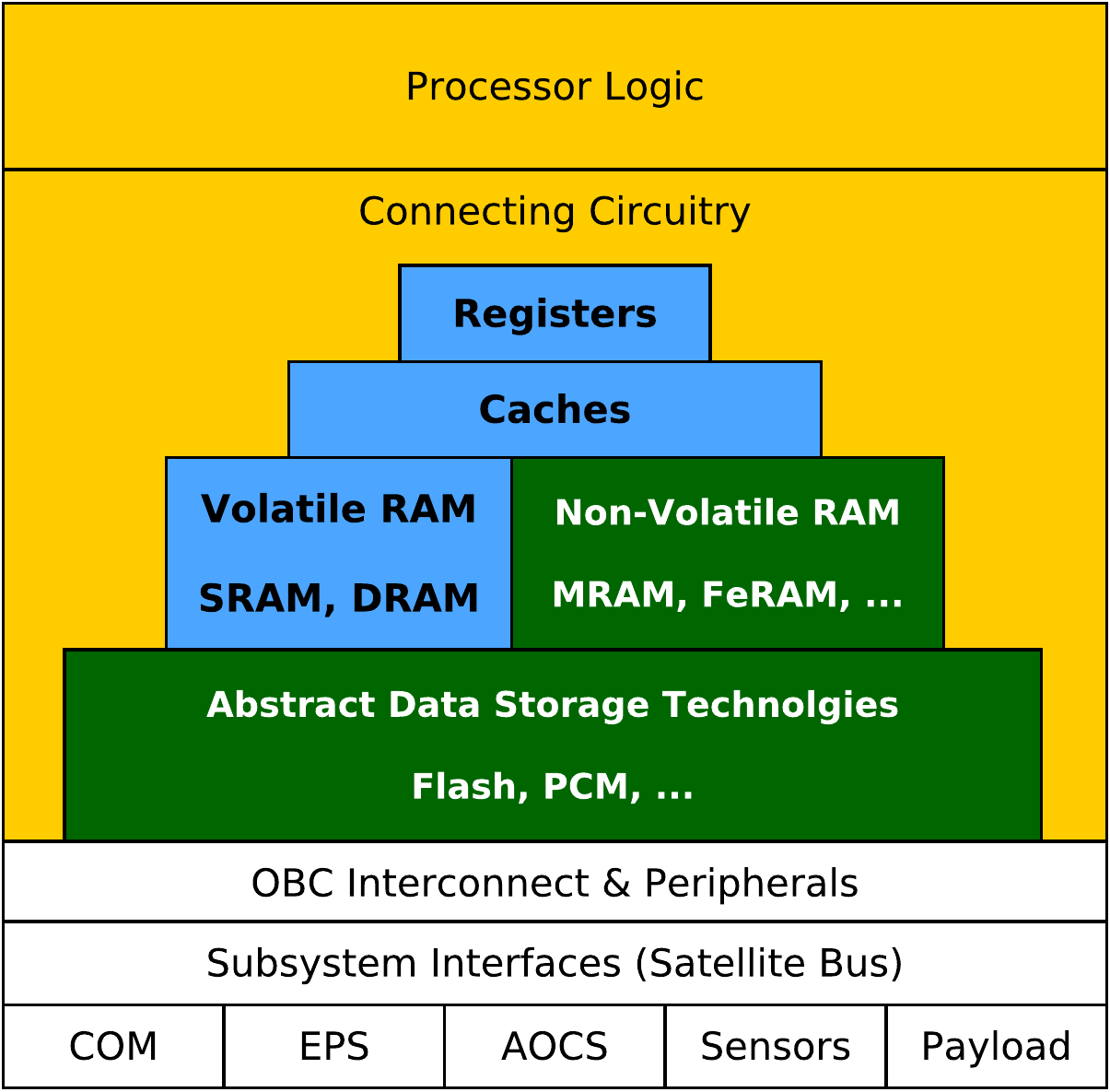}
		\caption{A component-wise view of a satellite OBC. Volatile memory (blue) and non-volatile memory (green) can well be protected using erasure coding. The presented multi-stage approach covers faults affecting processor logic (yellow).}
		\label{fig:vulnerability}
		\vspace{-9pt}
	\end{figure}

	The many challenges described in this section imply severe constraints to an OBC, but besides radiation, many can be tackled through wise engineering decisions and a smart system design.
	Radiation challenges OBC fault-coverage constantly and throughout a mission and affects all of an OBC's components depicted in Figure \ref{fig:vulnerability}.
	Radiation may corrupt the results of processed instructions, register content and data stored in caches, main memory, and non-volatile memory.
	Thus, not only application data is prone to data corruption, but faults can also affect all OS meta-data and kernel structures.
	Memory mainly suffers from bit-rot and malfunctions in controller logic, and for volatile memory, these can well be compensated for using error correcting codes (ECC) combined with error scrubbing.
	Non-volatile memory (green) also requires more powerful erasure coding systems, the basic notions of which also exist in latest-generation COTS flash memory based devices.
	Functional interrupts can also result in individual process cores or logic units (yellow) failing temporarily or permanently.
	Data can also be corrupted in transit, e.g. while being transferred or due to upsets in peripheral interface controllers.
	Hence, from developer perspective, to-be executed software and data can only be considered fault free if it resides exclusively in radiation-hard memory and radiation-hard processing logic throughout.
	As this is not the case with all but trivial processing logic, no part of an OS can be relied upon to be fault-free, and concepts requiring such an entity do not offer effective fault-coverage in the space environment.

	\section{Related Work}
	\label{sec:relatedwork}
	Traditionally, FT is enabled through circuit-, RTL-, core-, and OBC-level voting, which is costly to develop, difficult to validate, maintain, and slow to evolve \cite{reick2008fault, hijorth2015gr740, jackson2016implementation, iturbe2016triple, ludtke2014obc, gupta2015shakti}.
	Software takes no active part in fault-mitigation, as faults are suppressed at the circuit level, preventing the effective assessment of a processor's health.
	Circuit- and RTL-voting are effective for microcontrollers and very small SoCs, while core-level voting requires logic unavailable in COTS systems.
	Modern embedded COTS MPSoCs consume very little energy.
	But to achieve FT using hardware-side measures, arrays of synchronized high-frequency voters or core-lockstepping in hardware are necessary.
	As voting and core-level lockstepping at GigaHertz clock rates is non-trivial, it has been implemented only at considerably lower frequencies with non-COTS hardware \cite{gupta2015shakti, iturbe2016triple, decoursey2006non, pigno2011testbench}.
	In general, hardware-voting based MPSoC designs are static and non-adaptive, as the entire design's fault-coverage properties are highly chip specific \cite{pignol2006dmt}.
	All these components are single-vendor solutions, therefore implying walled-garden environments.
	FT MPSoCs for space use contain retrofitted TMRed single-core processors, e.g. \cite{hijorth2015gr740}, or are unique, experimental solutions for specific satellite missions \cite{hulme2004configurable, samson2011implementation}.
	In contrast to these solutions, modern MPSoCs also allow considerably more software design freedom due to the available compute resources, thereby reducing the required development time and complexity.
	For scientific instrumentation and low-priority CubeSat missions, COTS-based MPSoCs and FPGA-SoC-hybrids have been utilized, but these are not suitable for critical satellite control applications within miniaturized satellites \cite{iturbe2015use}.
	Ground-based FT applications do not consider the specific threat-scenario and application environment, physical constraints, and thermal design constraints \cite{marinella2013total, Schwank2013}.
	Instead, we propose to use software-side functionality to assure FT for conventional, non-fault tolerant processor cores.
	
	First concepts involving coarse-grained lockstepping are promising \cite{kretzschmar2016synchronization, dobel2014operating, shye2007using}, but do not address the specific challenges to FT in space \cite{dong2013colo}.
	FT using thread-level very-long-instruction word architectures \cite{sartor2017exploiting, anjam2013configurable} has also been explored, though the approach still requires pipeline-level voters in hardware.
	Most implement checkpoint \& rollback or restart, which makes them unsuitable for spacecraft command \& control applications \cite{hursey2007design}, others ignore fault-detection \cite{munk2015toward, zeng2016towards}, or require external, infallible fault detection entities with deep knowledge about application-intrinsics \cite{azad2016holistic} but no concept of how this could be obtained.
	Often, faults are assumed to be isolated, side-effect free and local to an application \cite{holler2015software} and/or transient \cite{dobel2014operating, shye2007using, munk2015toward}, which voids their effectiveness for space applications.
	Many prior concepts entail high performance- \cite{santangelo2013open}, resource-overhead \cite{missimer2014distributed, al2016fault}, or impose severe design constraints on applications and the OS \cite{kretzschmar2016synchronization, dobel2014operating}.
	To be effective in the space environment, an FT approach must be based upon forward-error-correction and the implementation complexity must be low, and must be suitable for general-purpose computing and impose little or no constraints on the application software.
	Changes to the OS infrastructure must be platform portable, code-wise localized, and individually verifiable.
	
	\cite{dobel2014operating, shye2007using, holler2015software} implement voting through OS invasive measures, can not handle multi-threaded applications and consider the OS and stored program code to be fault free.
	\cite{dong2013colo} requires no modifications to the application software whatsoever, but can only assure availability in a networked application architecture.
	An acceptance of these constraints does not allow for adequate FT in a space mission scenario, and thus we propose that application and OS instance must be able to fail arbitrarily without impacting the residual system.
	In this case, fault propagation between application instances also becomes a non-issue.
	Considerable research has been directed towards FT real-time scheduling and mixed critical software-FT systems, though only at a theoretical level \cite{malik2011adaptive, smiri2016fault, al2016four}.
	As a consequence, no implementable, software-driven FT concept for modern embedded- and mobile-market MPSoCs in space exists, creating a gap between the described prior research on software- and hardware-FT based implementations.
	
	\section{Bridging the Gap: Our Approach}
	\label{sec:overview}
	\noindent
	This approach consists of three fault-mitigation stages:
	\begin{itemize}[leftmargin=1.2cm]
		\item[\textbf{Stage 1}] is implemented entirely in software and provides fault-detection through coarse-grained lockstepping to enable self-testing, and can be implemented in COTS MPSoCs.
		\item[\textbf{Stage 2}] improves medium-term reliability, and enables long-term fault-coverage through FPGA reconfiguration and the use of alternative configuration variants.
		It utilizes Stage 1's fault detection capabilities.
		\item[\textbf{Stage 3}] extends the lifetime of a degraded OBC by utilizing mixed criticality to assure fault-coverage for high-criticality threads.
		It enables the OBC to automatically sacrifice performance or fault-coverage of lower-criticality threads in favor of higher-critical applications, thereby maintaining a stable core system.
	\end{itemize}
	The presented concept is flexible and the individual stages are modular, as Stage 2 or 3 can be omitted depending on the OBC and mission.
	Our approach is designed for generic COTS MPSoCs, as these are readily available in a variety of performance classes at low cost.
	The tiled architecture described in Section \ref{sec:platformarch} is optional but can be considered as an ideal platform.
	In MPSoCs without a tiled architecture, \emph{tile} can be substituted for \emph{processor core}, and the differences in fault coverage are discussed in Section \ref{sec:platformarch}.
	
	\subsubsection*{\textbf{Terminology}}
	Fault detection in our approach is based upon sets of tiles running two or more lockstepped copies of application threads.
	We refer to such a group of lockstepped threads as a \emph{thread group}.
	Timing-compatible thread groups can be combined and executed on the same set of tiles, and are then referred to as a \emph{tile group}.
	A tile group periodically executes a \emph{checkpoint routine}, which computes checksums for all active threads and compares them with the other tiles in the group (\emph{siblings}), thereby enabling a majority decision.
	The time between checkpoints (the \emph{checkpoint frequency}) is defined by the threads in a tile group and can be modified at runtime.
	All lockstepping-relevant information is stored in \emph{validation memory}, a tile-dedicated memory segment which is read-only accessible by tiles.
	
	\subsubsection*{\textbf{Application Requirements}}
	The OS only has to support interrupts, wake-up timers, and a multi-threading capable scheduler.
	To the best of our knowledge, such functionality is available in most widely-used RT- and general-purpose OS implementations.
	Virtual memory support is required to enable performance-efficient multi-threading.
	Furthermore virtual memory drastically simplifies thread-management, context switching, and thread isolation, benefiting overall fault-tolerance.
	
	The only requirement for applications is interruptibility at application-defined points in time, during which checkpoints can be executed.
	As there is no efficient, uniform approach to assess the health of threads, we rely upon applications assessing their own health-state.
	A thread provides four callback routines, which are executed during tile initialization and by the checkpoint handler:
	\begin{itemize}[leftmargin=.35cm]
		\item an \emph{initialization routine}, to be executed on all tiles at bootup;
		\item a \emph{checksum callback}, used to generate a checksum for comparison with siblings,
		\item a \emph{synchronization callback}, exposing all thread-state relevant data to synchronize a sibling with a tile group;
		This data can either be placed directly in the tile's validation memory, or as a reference to structures in main memory.
		\item and an \emph{update callback}, which is executed on a tile that needs to synchronize its state to a tile group.
	\end{itemize}
	Some of these callbacks may be omitted, e.g., for applications not requiring bootstrapping or with an already exposed state.
	The checksum computation and state (re)-synchronization are intentionally placed within the domain of the application developer.
	This enables decisions about an application state to be taken by the entity with the best knowledge of the individual thread and the means to determine which data is relevant to the system and application state, and must be preserved.

	Threads can be executed in an arbitrary order within a lockstep cycle as long as their state is equivalent during the next checkpoint.
	However, interrupting an active application at a random point in time is usually undesirable.
	We avoid thread-synchronization issues \cite{kretzschmar2016synchronization} by enabling the application developer to define comparison points where the application will yield control to the checkpoint handler.
	If an application requires real-time scheduling, the tightness of the RT guarantees depends upon the time required to execute these callbacks.
	Communication between thread-groups and tile-groups is of course possible and will remain reliable, as long as the receiving application is aware that it will receive multiple message replicas.
	To prevent faults from propagating through IPC channels, a thread can compare the received messages.
	
	\subsubsection*{\textbf{Limitations}}
	This approach guarantees system state consistency and control flow correctness after each checkpoint, and for all past checkpoint periods.
	It also assures computational correctness before the last checkpoint, but can not actively prevent faults from occurring during the ongoing checkpoint cycle.
	Thus, if one tile experiences a fault, incorrect results may be propagated outside the system, even though the damage caused to the OBC will be corrected during the next checkpoint, and system state consistency will be asserted.
	This limitation is inherent to coarse-grained lock-stepping concepts, but could be elevated at the thread-level somewhat using finer-grained event hooking, e.g., system-call hooking \cite{dobel2014operating}.
	However, this workaround requires in-deep modifications to the OS kernel and development toolchain, is thus non-portable and difficult to maintain, while still not solving the underlying conceptional limitation.
	
	Related research, however, does show that a solution at the system-design level is much better suited to prevent fault-propagation of transient faults between checkpoints using simple I/O voting \cite{dong2013colo}.
	Traditional hardware-FT approaches used in space computing are strong for assuring non-propagation of faults across interfaces using hardware-side voting, but can not protect the control-flow and system-state consistency efficiently.
	While the system state and system-level fault-tolerance are assured by Stage 1, and long-term system resilience are safeguarded in Stage 2 and 3, we can utilize simple I/O voting to prevent fault-propagation for tile groups.
	Performing I/O voting on interface is already common practice in space-borne computing, as considerable effort is put into providing interface redundancy aboard larger satellites.
	Small satellites, especially CubeSats, usually can not spare the additional energy, space and mass required for interface replication.
	For such spacecraft, I/O voting can be implemented on-chip using library IP.
		
	\section{Stage 1: Short-Term Fault Mitigation}
	\label{sec:stage1}
	Stage 1 offers software-controlled, thread-level, distributed majority voting and fine-grained fault logging within any COTS MPSoC with three or more processor cores.
	The objective of Stage 1 is to detect and correct faults at each checkpoint to assure computational correctness, control-flow consistency, and a consistent system state after each checkpoint.
	To do so, Stage 1 requires a processor guaranteeing sequential consistency.
	
	Instead of exerting direct control over the MPSoC, a supervisor can assure FT indirectly, as fault-coverage and control are distributed and enforced by the tiles themselves.
	In consequence, the supervisor does not require any knowledge about the executed application threads, an individual tile's state, or other OBC intrinsics.
	The thread group assignment within an MPSoC can be reconfigured freely at runtime to implement different voting configurations.
	Thus, the described approach can exploit parallelization to improve reliability, throughput, or minimize power consumption, thereby allowing the system to adapt to multi-phased missions with varying performance requirements.
	
	\vspace{-5pt}
	\subsection{Thread-Based Self-Testing}
	The program flow of this stage is depicted in Figure \ref{fig:logic} and described below.
	It can be implemented within an existing scheduler and an interrupt service routine (ISR).
	A practical example for tile fault handling and recovery, and an overview over how the supervisor interacts with the system are provided at the end of this section.

	\subsubsection*{\textbf{Bootup \& Initialization}}
\begin{figure}[!t]
		\vspace{-6pt}
		\centering
		\includegraphics[width=0.9\linewidth]{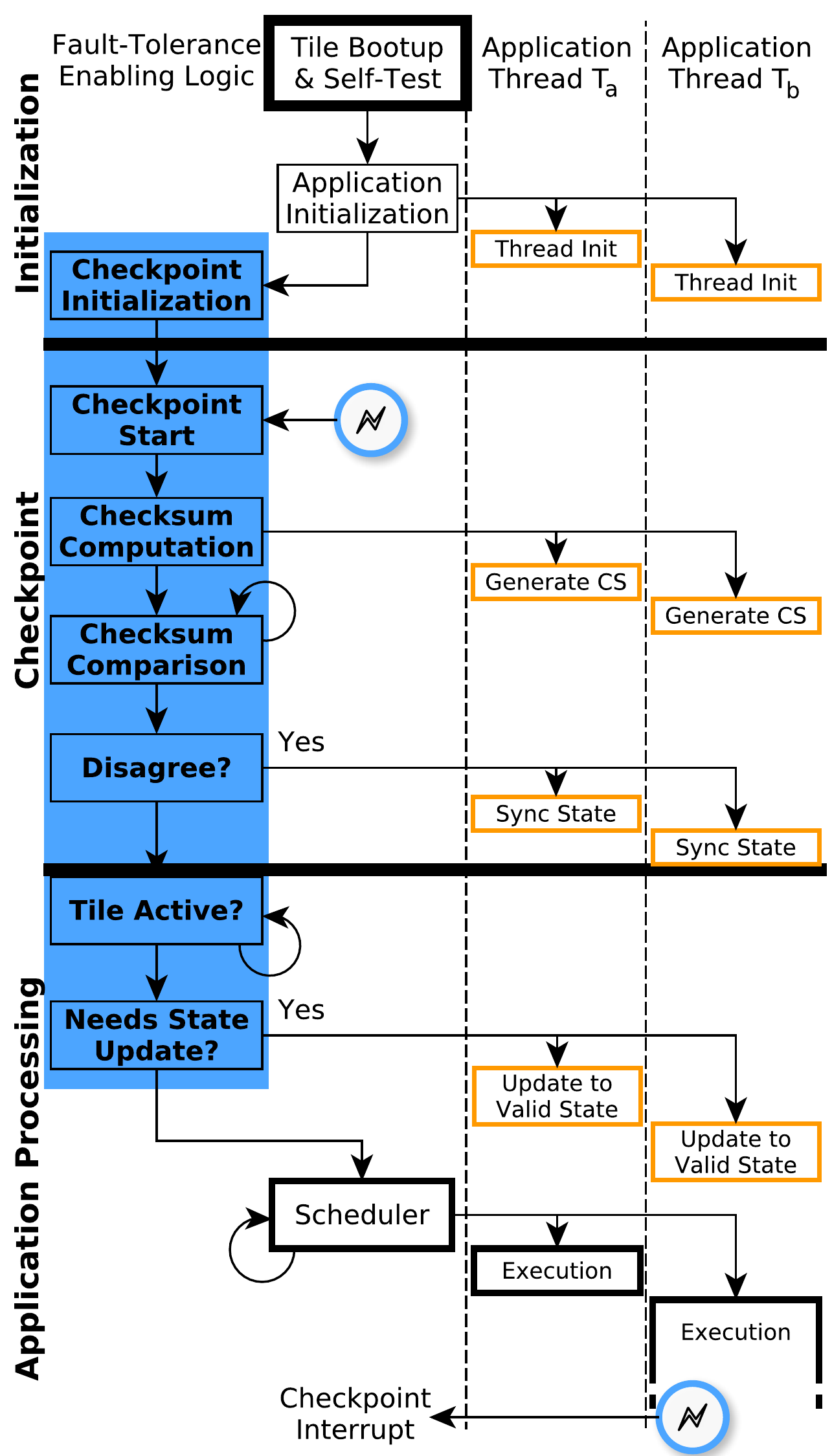}
		\vspace{-9pt}
		\caption{The execution cycle of a tile during Stage 1.
			All code necessary for implementation is highlighted in blue, callbacks in yellow.}
		\label{fig:logic}
		\vspace{-12pt}
	\end{figure}

	After bootup, a tile first executes basic self-test functionality to assure integrity of tile-local IP-cores and memory.
	Each thread's initialization routine is executed on all tiles to allow more rapid state-update in case a new thread-group is added to a tile.
	When being assigned to a tile, a thread will register its desired checkpoint frequency and its checksum, synchronization and update callback routines.
	After the threads have been initialized, each tile will set a periodic timer to initiate checkpoints.
	As depicted in Figure \ref{fig:logic}, a tile will execute its first checkpoint immediately after the MPSoC has been fully rebooted, to assure that application and OS initialization were successful.
	If only this individual tile was rebooted, it can thus return to the spare tile pool to replace a faulty core in the future.
	
	\subsubsection*{\textbf{Checkpoint Start}}
	A checkpoint is triggered by a timer interrupt or externally by the supervisor.
	A thread can delay a checkpoint until it has reached a viable state for checksum comparison by disabling interrupts, thereby deferring interrupt processing.
	The checkpoint ISR saves the existing system state, loads the actual checkpoint handler, performs a context switch to kernel mode, and invokes the checkpoint handler.
	
	\subsubsection*{\textbf{Checksum Computation}}
	\label{sec:CScomp}
	The checkpoint handler invokes each active thread's checksum callback scheduled for checking.
	As not all threads in a tile group require the same checking frequencies, not all active threads will be validated during each checkpoint.
	This checksum callback returns a representation of the application thread's internal state as checksum or hash generated from thread-private variables and other internal application state.
	The checksum format is compile-time defined, and must be chosen based on FT needs.
	The algorithm used to generate this checksum is up to the application developer.
	Each checksum is stored in the tile's local validation memory and thereby exposed to the other tiles.
	If no checkpoint routine can be provided, a checksum is computed by the checkpoint handler for an application-defined memory range.
	This memory range can be utilized by the application to deposit state-relevant data passively, e.g., through linker scripts or pre-processor macros.
	A non-continuously running application can also deposit its results in validation memory or return a checksum upon exit.
	
	Prior concepts required deep modifications to the OS to allow a proprietary central health-management entity to retrieve this information directly \cite{kretzschmar2016synchronization, munk2015toward}, or utilized no application-internal information \cite{dong2013colo, shye2007using, al2016fault}.
	Instead, this approach enables us to utilize application-intrinsics to assess the health-state of the system, without requiring any knowledge on the applications.
	The time required to generate checksums can be minimized by adapting the application code, e.g., by retaining computational by-products which would usually be discarded.
	
	\subsubsection*{\textbf{Checksum Comparison}}
	\label{sec:CScomparison}
	
	Once all checksum callbacks have been executed, a tile will monitor its group members' validation memory segments until another tile is ready for comparison.
	It will do so until it has compared its checksums with all siblings, or the system designer's tile-group deadline expired.
	Tiles will usually begin comparing its checksums with siblings immediately or wait only briefly, as delays are mainly induced due to varying memory latencies or malfunctions.
	If it detects a checksum mismatch or a sibling violated the deadline, the tile will stop comparing checksums and report disagreement with that tile to the supervisor.
	
	\subsubsection*{\textbf{Thread Disagreement \& State Propagation}}
	If a tile detected a checksum mismatch, it executes the synchronization callback routines for all threads in the affected tile group.
	This callback can be omitted if all state-relevant data is already in validation memory, e.g. for non-continuous running applications.
	The checkpoint routine will adjust the checkpoint's timer if a new thread group was added to the tile group, and return control to the scheduler.

	\subsubsection*{\textbf{State Update and Thread Execution}}
	The scheduler will check three conditions during regular operation: if any thread-group is active, the tile was newly added to a tile group, or requires an update.
	Idle tiles sleep until the next checkpoint and can be woken up by the supervisor to reduce energy consumption and fault-potential.
	In case a tile must update a thread-group's state from a sibling, the relevant update callback will be executed for each thread.
	Tiles that have detected disagreement with one of their siblings will delay execution for a tile-group-wide grace period, to allow a sibling to retrieve a state-copy from validation memory.
	Once a tile has updated its state using a sibling's data, application processing continues.
	The other tile group members will also wake up after the grace period and continue executing threads.
	This concludes the lockstep cycle.
	
	\begin{figure}[!b]
		\vspace{-18pt}
		\centering
		\includegraphics[width=0.95\linewidth]{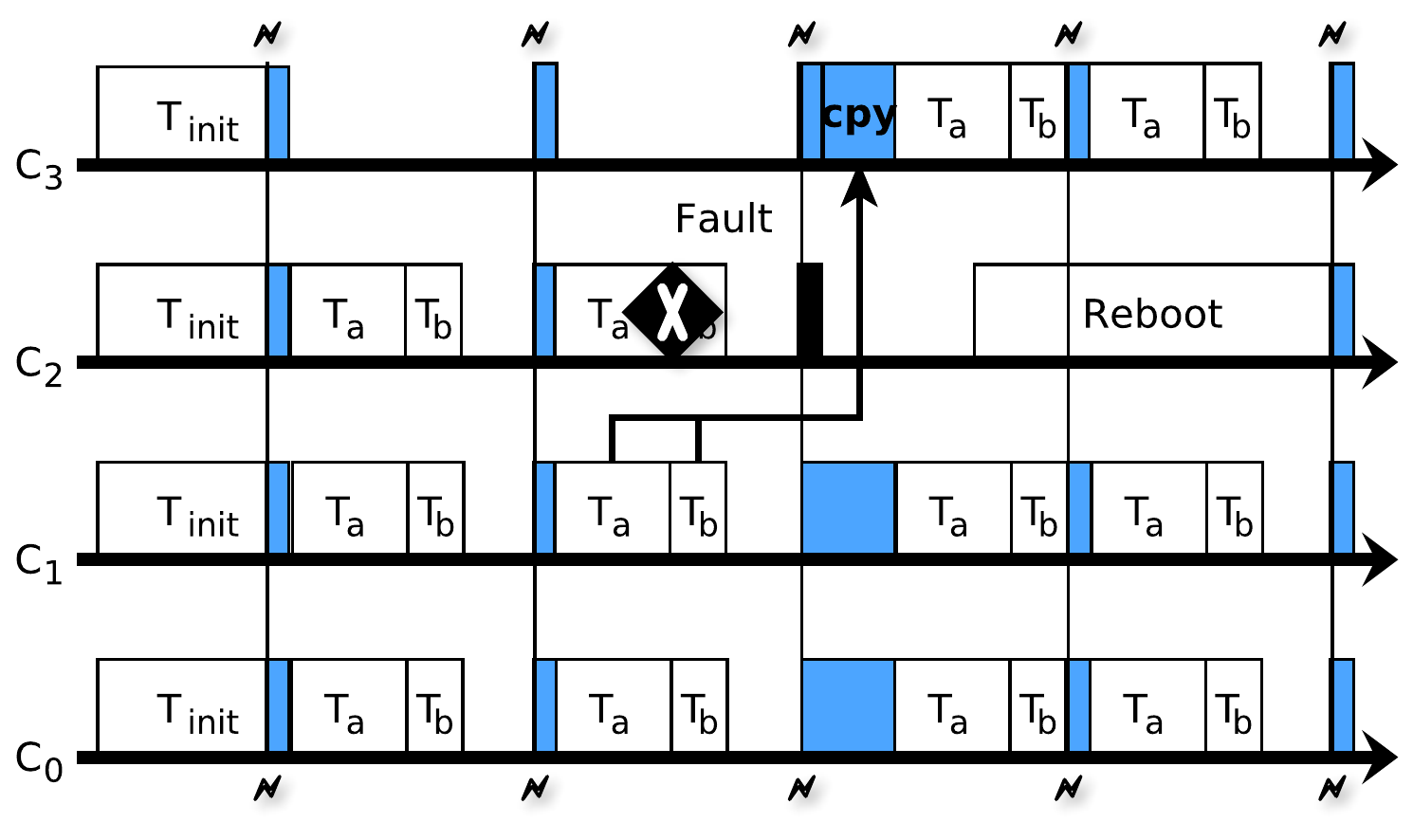}
		\vspace{-15pt}
		\caption{Tile initialization and a complete Stage 1 lockstep cycle.}
		\label{fig:ThreadRecovery}
		\vspace{-2pt}
	\end{figure}

	\vspace{-3pt}
	\subsection{A Practical Example}
	Figure \ref{fig:ThreadRecovery} depicts a quad-core MPSoC with a single tile group and three members.
	A fault has occurred during the second lockstep cycle on tile $C_2$, which is subsequently replaced with the idle tile $C_3$.
	$C_3$ must retrieve a copy of the state of its threads $T_{a}$ and $T_{b}$ from another valid sibling.
	The replaced tile, $C_2$, can subsequently be tested for permanent defects by the OS and the supervisor.
	
	\vspace{-3pt}
	\subsection{Checkpoint-Frequency, Timing \& Real-Time Capabilities}
	The level of fault-coverage is mainly dependent on the checkpoint frequency.
	During a checkpoint, the computationally most costly operations are the application checksum callbacks, the synchronization callbacks and a new tile's update callback.
	Each of these operations involves a context switch and may imply a varying level of data being read or written.
	Thus, the performance overhead and fault-tolerance capabilities are mainly based upon actual applications checked, as this actual checkpoint handler code is rather trivial.
	In general, a higher checkpoint frequency implies more time will be spent in checkpoints, more fine-grained fault-detection are possible, thus better fault-coverage.
	
	In our implementation, interrupts are deferred during a checkpoint, thus applications are not serviced and will not process I/O, thereby affecting the level of real-time capabilities the MPSoC can offer.
	However, though this can be worked around using a more elaborate interrupt handling concept, e.g., using interrupt prioritization or filtering.
	Real-time capabilities are thus directly dependent on the MPSoC, and application implementation characteristics, with the OS infrastructure playing a minor role.
	For complex applications with a large state, a lower checkpoint frequency however also implies a larger difference in state.
	Hence, more data must be copied between tiles to achieve thread-synchronization requiring additional time.
	Thus, a larger state also requires more time for execution, potentially more complex data structures, thereby implying longer synchronization- and update-callback.
	
	Overall, the performance of OBCs executing less complex applications with little state will improve with lower checking frequencies.
	For such OBCs, more checkpoints imply more computational overhead.
	With more complex applications, there is considerable optimization potential to find a sweet-spot between checkpoint frequency and application-state size.
	However, performance is strongly dependent assuring that high-quality callback-routines are provided by the application developer.

	\subsection{Supervision}
	\label{sec:supervision}
\begin{figure}[!t]
		\centering
		\includegraphics[width=1.03\linewidth]{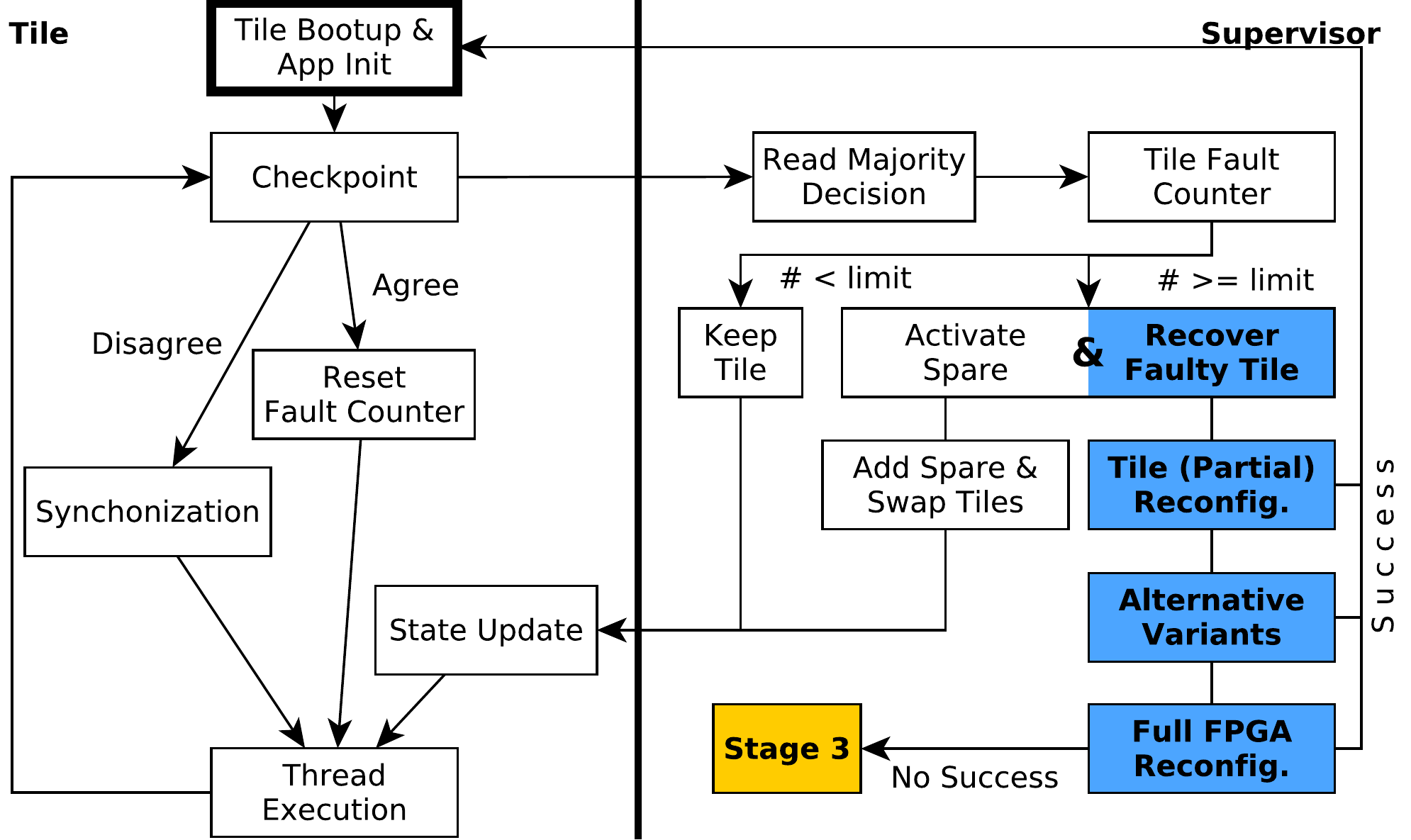}
		\vspace{-9pt}
		\caption{A tile's and supervisor's program-flow and their interactions.
			Stage 1, 2 and 3 logic are indicated in white, blue and yellow respectively.}
		\label{fig:supervision}
		\vspace{-9pt}
	\end{figure}

	The supervisor is connected to the MPSoC through a multiplexed bus-interface, where each line signals agreement with another tile.
	Finer-grained disagreement reporting does not significantly improve fault-coverage and constrains scalability of the MPSoC.
	As depicted in Figure \ref{fig:supervision}, the supervisor only reacts to disagreement between tiles, otherwise remaining passive.
	It maintains a fault-counter for each tile, and acts as a system-reset inducing watchdog timer for the MPSoC.
	To resolve transient faults within a tile, it increments the fault counter and induces a state update through a low-level debug interface.
	After repeated faults, the supervisor will replace the tile by adjusting the thread-mapping of a spare tile, activating it, and rebooting the faulty tile.
	In case a system developer indicated threshold is exceeded, the disagreeing tile is assumed permanently defunct and not re-used as a spare.
	Stage 1 alone can not reclaim defective tiles beyond programmatically avoiding the use of defective peripherals, memory pages or processor functionality.
	Thus, Stage 2 will attempt to repair tiles to prevent resource exhaustion.
	
	In contrast to existing FT solutions, faults can be reported by each tile individually, because fault detection is decentralized.
	As this functionality is implemented at the kernel level, we can utilize the OS's powerful logging and diagnostics facilities, instead of relying upon the supervisor to provide a minimal useful level of logging.
	Diagnostics can thus be enriched with application-level information.
	Thereby, defect assessment accuracy can be improved compared to prior FT-approaches, enabling more sophisticated debugging without requiring live-interaction.
	
	Our approach enables lockstepping frequencies far below the KiloHertz range, thus the supervisor will not be a bottleneck.
	Therefore, high-performance MPSoCs can be well supervised using pre-existing discrete COTS supervisors.
	COTS MPSoCs will utilize an external supervisor, while ASIC, FPGA and FPGA-SoC-hybrid based MPSoCs can implement this functionality in reconfigurable logic.
	An off-chip supervisor can be used for active tile health-management and FPGA reconfiguration, enabling the use of FPGA reconfiguration.
	See \cite{fuchs2017performance} for further details on MPSoC to Supervisor communication.
	
	\section{Stage 2: Tile Repair \& Recovery}
	\label{sec:stage2}
	The previous stage can compensate faults as long as healthy tiles are available to replace defective tiles.
	In all existing hardware-side FT implementations, resource exhaustion is mitigated through over-provisioning (adding more spares).
	Over-provisioning of tiles naturally is inefficient and curtails system scalability, but is certain due to the static, unchangeable nature of existing ASIC based solutions.
	This will inevitably result in resource exhaustion, and has not been solved in prior work.
	Stage 2 is designed to perform active tile health management and test, repair, validate and recover faulty tiles, thereby tackling this fundamental limitation.
	In FPGA-based systems transient faults can corrupt the stored configuration of programmed logic, thus induce permanent effects within the running configuration \cite{azimi2016prediction, zhang2016aging}.
	However, even if a logic cell is damaged permanently the residual highly-redundant FPGA fabric will remain intact and can be re-purposed \cite{siegle2015mitigation}.
	It could be repaired with differently routed, functionally equivalent configurations.
	
	The main issue preventing prior research from utilizing FPGA reconfiguration to increase FT of general purpose computing architectures is a lack of non-invasive, flexible circuit level fault detection.
	As efficient fault-detection is an unresolved issue and periodic configuration scrubbing is slow, Stage 2 relies upon fault-detection by Stage 1.
	If a tile was replaced by a spare, the supervisor's Stage 2 logic recovers tiles using partial reconfiguration, mapping a tile to one of multiple partitions.
	Once reconfiguration is complete, the supervisor validates the relevant partitions to detect permanent damage to the FPGA fabric.
	Assuming a tiled MPSoC architecture (see Section \ref{sec:platformarch}) is used, tiles are self-contained by design.
	Thus, reconfiguration of just one tile will not impact the other tiles and allow the OBC to recover a tile in the background.
	If reprogramming was unsuccessful or fabric-level faults persist, the supervisor will repeat the previous step with differently routed configuration variants.
	Partially defective logic cells can be re-purposed, while other cells can be avoided entirely, if no other usage is possible.
	Other elements of the FPGA fabric can be treated equivalently.
	As a final measure, faults within shared logic can be resolved using full reconfiguration, briefly halting the MPSoC.
	
	Stage 2 can also test different on-chip memories, the processor cores, and peripheral controllers through external interconnect access ports (e.g. an AXI-bridge).
	If the OBC is implemented on an ASIC or with a COTS MPSoC, a widely available low-level debug and testing interface such as JTAG can be utilized for the same purpose.
	For further details on how this functionality can be implemented, see \cite{fuchs2016enhancing}.
	
	If a defunct tile can not be repaired through automated reconfiguration, additional diagnostic information can be used for further analysis.
	The operator can utilize this information to conduct fault analysis on the ground, to craft a suitable replacement configuration to avoid these areas.
	Of course, this implies extreme development effort but for many higher-priority space missions, the loss of a spacecraft may be more costly than the engineering costs for saving the mission.
	If both partial- and full-reconfiguration are unsuccessful and all spare resources have been exhausted, the fault is escalated to Stage 3.
	
	\section{Stage 3: Applied Mixed Criticality}
	\label{sec:stage3}
	Stage 3 utilizes thread-level mixed criticality to extend an OBC's lifetime once the previous stages have depleted all spare resources.
	Its primary objective is to autonomously maintain system stability of an aged or degraded OBC at short notice to avert loss-of-mission and loss-of-subsystem, even if an OBC approaches the end of its lifetime.
	The operator can then define a more resource conserving satellite operations schedule, sacrifice link capacity, or on-board storage space.
	Thus, dependability for high-criticality threads can be maintained by reducing compute performance, throughput, or increasing latency of lower-criticality applications.
	
	The criticality of applications executed on an OBC can be differentiated by the importance of the controlled subsystem or relevance for commandeering the spacecraft.
	Performance degradation or even a loss of lower-criticality tasks aboard a satellite is in general preferable to a loss of system stability for key applications.
	As thread groups can be added and removed from tile groups, and multiple tile groups can coexist in the same MPSoC, individual threads can also be migrated between tile groups \cite{zeng2016towards}.
	Furthermore, the checkpoint frequency of a tile group can be reduced to increase a tile's computational capacity, or it can cease servicing low-priority interfaces.
	
	The supervision logic is extended to reallocate thread-groups across the system based upon the thread's priority.
	Hence, if Stage 2 failed to reconfigure the OBC, the supervisor can generate new tile-group assignments for threads with high priority and will attempt to retain existing assignments.
	Eventually, all healthy tiles will be saturated with threads, and no further assignments will be possible.
	Then, it can either allocate more mappings, providing lower-priority threads with less processing time to maintain availability, reduce the checking frequency, or leave them inactive.
	The OBC developer can decide at design time, which applications would benefit most from continuous operation with reduced performance or reliability, and which can be forgone.

\begin{figure}[!t]
		\centering
		\includegraphics[width=1\linewidth]{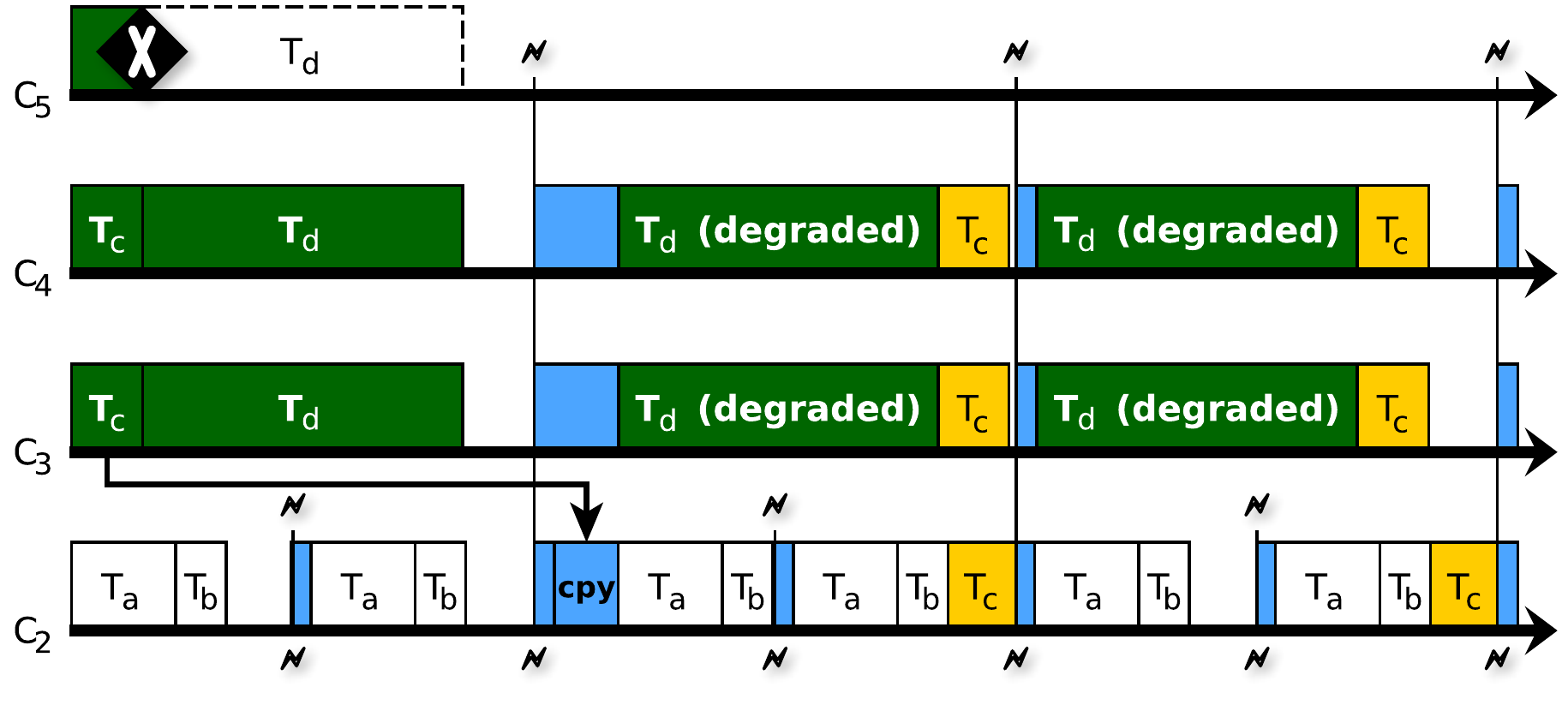}
		\vspace{-24pt}
		\caption{If no healthy spare tiles are available, the Stage 3 can split defunct tile groups and uphold FT guarantees for high-criticality threads.
			The necessary adjustment to the checkpoint frequency on tile 2 is omitted for simplicity.}
		\label{fig:ThreadMigration}
		\vspace{-12pt}
	\end{figure}

	In Figure \ref{fig:ThreadMigration}, initially two tile groups are executed on one MPSoC with 6 tiles.
	The green tile group consisting of a computationally expensive low-criticality application $T_{d}$ and a shorter but more important thread $T_{c}$.
	Tile 2 is member of another group, and has sufficient spare capacity to accommodate $T_{c}$, but not $T_{d}$.
	As no more spare tiles are available, the lower-criticality task $T_{d}$ remains degraded, and can only detect but not correct subsequent faults.
	$T_{c}$ is migrated to a separate, new tile group and executed on tiles 2 -- 4, thereby maintaining strong FT.

	\section{Platform Architecture}
	\label{sec:platformarch}
	Our multi-stage FT-approach is in principle platform independent and can be implemented within any multi-threading capable OS supporting interrupts and timers.
	For most COTS-MPSoC based nanosatellites in a LEO orbit, stage 1-3 alone offer sufficient fault-coverage.
	Aboard such spacecraft, MPSoC interfaces are either unprotected or protected programmatically and outside the MPSoC (e.g. using EDAC chips or by resolving SEFIs through power cycling).
	Aboard larger, more critical spacecraft such faults can not be accepted, and OBC interfaces are usually implemented redundantly at great effort.
	This redundancy is inherent to our approach with tiled architectures, and we developed an MPSoC platform capable of surviving the loss of peripheral devices and permanent, non-resolvable defects in interfaces.
	
	\subsection{Architecture Overview}
	
	This MPSoC can completely be implemented in full using library IP available with standard industry FPGA or ASIC design tools without custom FT components.
	We have implemented our MPSoC prototype with Xilinx Vivado standard IP, AXI Interconnects, for low-tier ARM Cortex-A processor cores to be provided by one of our industrial partners.
	For common space applications, size-optimized cores such as the Cortex-A32, -A35 and A5 offer an excellent balance between performance, universal platform support and logic utilization.
	The architecture minimizes shared logic, compartmentalizes tiles, and offers a clearly defined access channel between tiles for sharing checkpoint-results and application-state.
	We are aware that most miniaturized satellites do not require such a high degree of fault-coverage, and often can not afford the added hardware complexity and development effort.

	\begin{figure}[!b]
		\vspace{-9pt}
		\centering
		\includegraphics[width=0.8\linewidth]{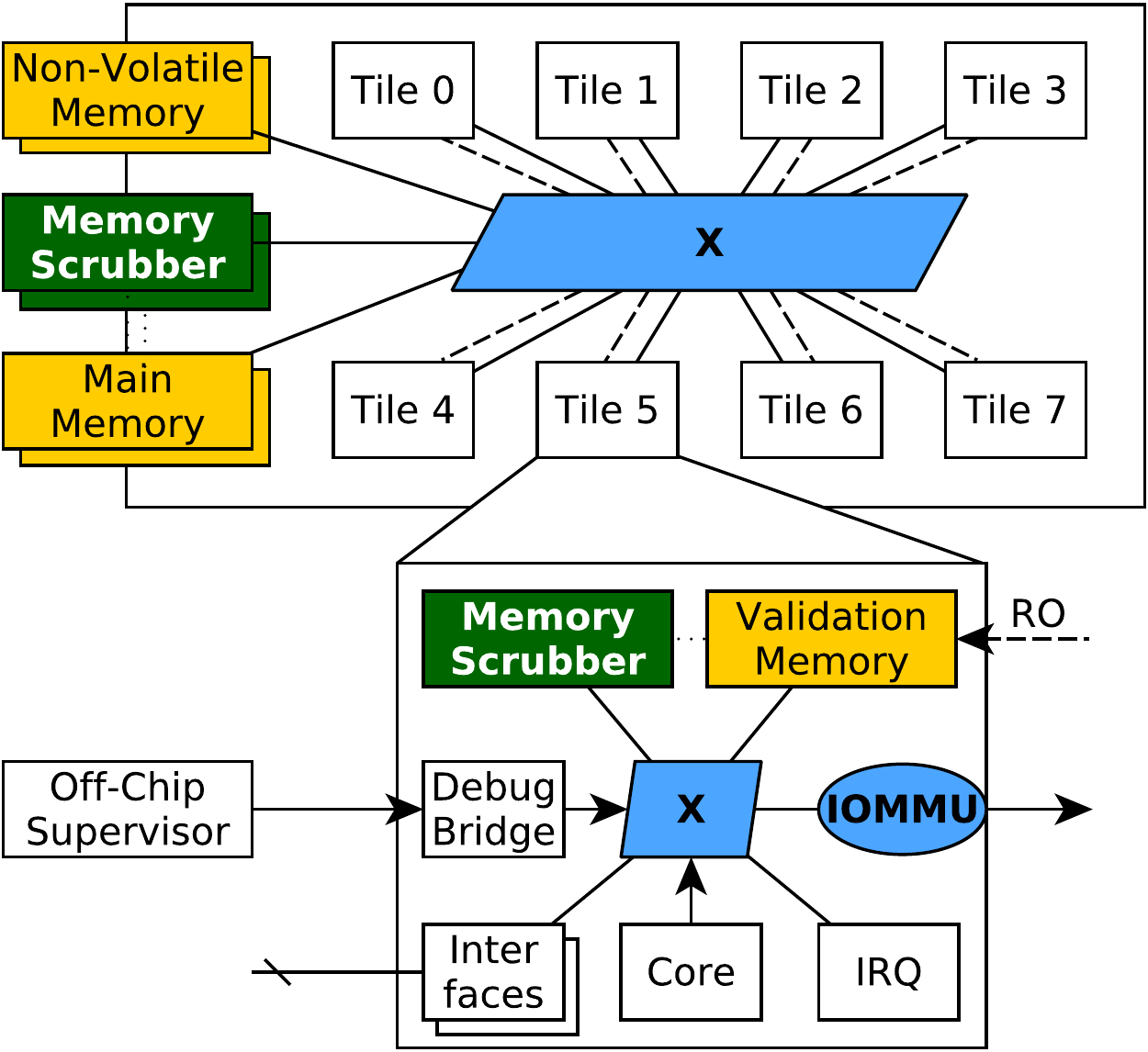}
		\vspace{-3pt}
		\caption{A simplified representation of the presented MPSoC with memory controllers highlighted in yellow, scrubbers in green, and interconnect in blue. A dedicated interface on each tile allows supervisor access.}
		\label{fig:tile}
		\vspace{-2pt}
	\end{figure}

	The design depicted in Figure \ref{fig:tile} follows a tiled architecture and is implemented within an FPGA to counter resource exhaustion when mitigating faults in Stage 2.
	It utilizes simple redundancy to compensate for SEFIs, but does not contain radiation-hard or FT processor cores or custom logic.
	Each tile is equipped with a processor core, an interrupt controller (IRQ in the figure), a dedicated on-chip memory slice used as validation memory, and several peripheral interfaces through the local interconnect.
	Tiles are connected through an I/O memory management unit (IOMMU) and a global interconnect to main- and non-volatile memory.
	They can not access the local interconnect of other tiles to prevent interference and minimize shared logic.
	This tiled architecture benefits from partial reconfiguration, as tiles can be placed strategically on an FPGA's fabric along partition borders.
	Our approach and this architecture support multi-FPGA and -ASIC MPSoCs without adaptation, thereby improving scalability and resilience against FPGA-level SEFIs.
	
	The ECC-protected dual-port validation memory in each tile holds the current tile-status, thread assignments, as well as the checksums and state information.
	One interface is connected to the tile's local interconnect, while the second port is read-only accessible via the global interconnect.
	The validation memory is inherently redundant, as threads are executed on at least two tiles.
	The shared main memory is redundant to safeguard from SEFIs affecting the tile-shared interface.
	Both instances are ECC protected and connected to the global interconnect.
	The main memory is split into several segments: each tile has write-access to its own segment, and can read the global shared code segment.
	ECC-fault syndrome interrupts for main memory are handled by the supervisor.
	We perform error-scrubbing on these memories to avoid accumulating bit-flips due to transient and permanent faults.
	The scrubbing frequency should be set depending on the actually used memory technology, production node and mission parameters.
	Non-volatile memory is implemented redundantly as well.
	Our prototype is designed to utilize radiation immune MRAM and PCM \cite{marinella2013total} to enable advanced FT data storage concepts \cite{fuchs2015ftrfs, RAID5FTLinHardware}.
	Each tile's main memory segment, validation memory, and non-volatile memory are mapped to the same tile-local address ranges.
	At the thread-level, the address-space in each tile is thus identical, making application and OS code location independent and allowing tiles to share binaries.
	Further implementation details are available in \cite{fuchs2017performance}.
	
	\subsection{FPGA Implementation \& Utilization}
	We also have developed a community reproducible MPSoC design based on the previously described architecture utilizing exclusively library-IP.
	Instead of ARM cores, this 4-tile demonstration design includes Xilinx MicroBlaze processor cores, as these are more available to the general public.
	It targets standard FPGA development boards and is equipped with a single shared DDR4 main memory controller, and 2MB on-chip BRAM program memory.
	This reduced design was implemented successfully using the Xilinx Vivado Design Suite and Stage 1 was implemented using FreeRTOS and using the Xilinx SDK toolchain.
	Each tile is outfitted with data and instruction caches, an interrupt controller, a UART interface, validation memory and an additional local memory for storing tile-private information, and a GPIO controller to signal agreement between tiles.
	All tile-local memories are equipped with ECC, as this increases logic size of the relevant memory controllers, and includes two additional interrupts for each connected memory.
	We could achieved full timing closure at 250MHz core frequency on VCU118 and KCU116 development kits, though the clock frequency was selected to achieve a simple design, not an efficient or fast one.
	If additional time was invested into timing optimization and clocking, the clock speed can be drastically increased.
	Additional information regarding the tile and SoC layout are available in \cite{fuchs2017performance}.
	
\begin{figure}[!t]
		\centering
		\includegraphics[width=0.89\linewidth]{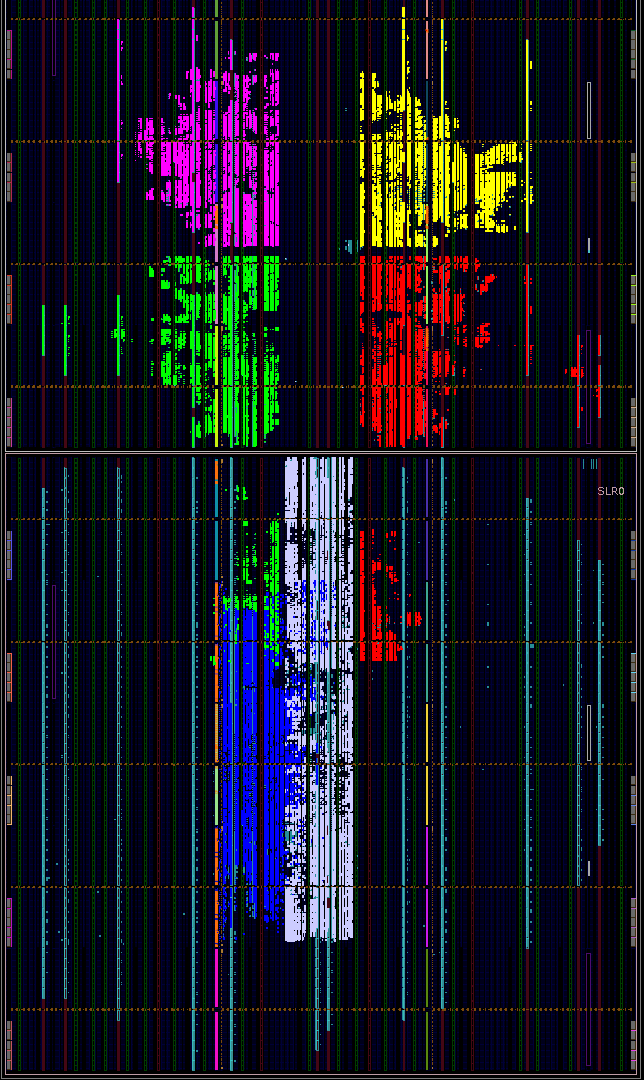}
		\vspace{-3pt}
		\caption{Logic placement of the demo-MPSoC on a VCU118 development board running 4 Tiles: green, red, yellow, pink; Global Interconnect: white; Xilinx DDR4 controller: blue; Program Memory: teal.}
		\label{fig:usagevisual}
		\vspace{-15pt}
	\end{figure}

	Fabric utilization based upon the Xilinx Virtex VCU118 Development Kit is depicted in Figure \ref{fig:usagevisual}.
	Due to the use of on-chip program memory and the DDR4 memory controller, BRAM utilization is inflated compared to the MPSoC described previously.
	Resource utilization is indicated in Table \ref{tab:usage}, with more details given in \cite{fuchs2017performance}.
	Stage 2 and 3 do not require additional FPGA logic.
	
		\begin{table}[!b]
		\vspace{-12pt}
		\renewcommand{\arraystretch}{1.3}
		\centering
		\begin{tabular}{l|r|r|>{\bfseries}r}
			Resource & Utilization & Available & Utilization \% \\
			\hline
			LUT	& 68,705	& 1,182,240	& 5.81\% \\
			LUTRAM	& 9,235	& 591,840	& 1.56\% \\
			FF	& 92,536	& 2,364,480	& 3.91\% \\
			BRAM	& 810	& 2,160	& 37.48\% \\
			DSP	& 27	& 6,840	& 0.40\% \\
			IO	& 163	& 832	& 19.59\% \\
			BUFG	& 17	& 1,800	& 0.94\% \\
			MMCM	& 6	& 30	& 20.00\%
		\end{tabular}
		\vspace{9pt}
		\caption{Resource utilization of the 4-tile demonstration MPSoC on a Xilinx VCU118 development board.
			The on-chip program memory and DDR4 memory controller disproportionately inflate BRAM utilization.}
		\label{tab:usage}
		\vspace{-12pt}
	\end{table}

	This design's very low logic usage shows that the architecture itself can be scaled to 8 and more tiles comfortably, and most current-generation FPGAs offer an abundance of unused resources for Stage 2.
	With current-generation FPGA platforms, Stage 2 will thus not only be able to recover defective tiles using spare resources, but could even place multiple tiles as cold or hot spares.
	The Microblaze cores utilized here for demonstration purposes can directly be replaced with drastically more complex processor cores, assuming the necessary peripheral IP is added as well (e.g. an ARM GIC instead of the MicroBlaze Interrupt Controller).
	
	\vspace{-6pt}
	\section{Discussion \& Outlook}
	\label{sec:discussion}
	The reliability of each individual tile's voting decision can be weak, and an individual tile can report false (dis)agreement with its siblings.
	Our approach takes into account that any software or hardware component associated within a tile can fail arbitrarily.
	Such failure is mitigated through a distributed decision, which is taken based on each tile's perspective of its siblings.
	Thus, this approach does not require the checksum logic to compute correctly, and we assume that faults may occur at any time during the lifetime of a tile.
	As tile groups usually consist of three or more tiles, the likelihood of false-disagreements or non-reported disagreement is insignificant.
	To mask such a fault, multiple faults would have to coincide in a majority of tiles within the same tile group during a single checking period and induce the same fault.
	The probability for such an event is extremely low, except at very high radiation levels.
	Even in such situations, such faults would be detected after the subsequent checkpoint with near certainty.
	
	Prior research proves the conceptual effectiveness of thread-based FT \cite{shye2007using, iturbe2016triple} and software-based FT combined with simple I/O voting \cite{dong2013colo}.
	Also, the detailed FT capabilities of a platform utilizing our approach are influenced by the actually used FPGA, ASIC or COTS-MPSoC design.
	These imply mainly design decisions and a varying acceptance of single-points-of-failure.
	Schedulability, timing conformity, and deadlock-avoidance have been extensively researched in literature, e.g., in \cite{missimer2014distributed}.
	Thus, what remains to be shown is the runtime performance overhead induced by the presented approach, as the main objective of our research is to enable the efficient use of high-performance mobile-market COTS MPSoCs within satellite computers.
	To achieve worst-case performance estimations, we developed a naive, unoptimized implementation of the Stage 1 of our approach, as the others do not affect the runtime performance of the MPSoC. 
	This naive implementation shows a median-best performance degradation of 9\% and median-worst degradation of 26\% on tiles with a single processor core.
	Further information on the conducted tests is available in \cite{fuchs2017performance}, as well as performance measurements for 6 different application scenarios modelled after the NASA/James Webb Space Telescope's Mid-Infrared Instrument (MIRI) \cite{ressler2015mid}.
	
	As prior thread-level FT implementations \cite{dobel2014operating, shye2007using, holler2015software} are based upon fundamentally different concepts, only address transient faults within a very limited scope, and are deeply embedded into proprietary OS, their fault-coverage and performance can not be directly compared.
	However, the measured performance overhead does fall within the same range as measured in \cite{dobel2014operating}, and we also observe comparable average-case performance.
	To put these measurements into context, even a 50\% slowdown on modern MPSoCs will offer a factor-of-5 performance increase over state-of-the-art radiation-hardened processor designs, thereby showing a favourable cost-vs-benefit trade-off.
	
	\section{Conclusions}
	\label{sec:conclusions}
	In this contribution, we present the first practical and integral multi-stage approach to fault-tolerant (FT) general purpose computing for spaceflight use.
	The approach explicitly does not utilize radiation-hardened or hardware-FT processor cores and utilizes no central MPSoC-internal voting logic.
	It can thus be implemented within COTS MPSoCs or alternatively entirely with non-FT, standard library IP-cores available in FPGA or ASIC design software.
	In contrast to prior research, the presented approach considers the full and realistic fault-model for space computing, and operates within real-world constraints.
	The approach does not require failure-free components within an MPSoC or in the OS, and does not leave conceptual gaps, e.g., regarding fault detection and recovery.
	It is not based upon traditional radiation-hardened processor cores and does not achieve fault-tolerance through hardware-measures.
	
	We showed that our approach is programmatically simple and requires little custom code, which can also be implemented in most pre-existing multi-threading capable OS.
	Faults can be detected and mitigated using application provided routines, enabling decisions about an application's integrity to be taken by the application developers themselves.
	As a consequence, the system designer no longer must struggle to assess the health of each individual application's state, and instead can focus on determining an optimal solution to problems at hand.
	It allows flexible fault-detection, mitigation and recovery within COTS MPSoCs, laying the foundations for FT computing aboard miniaturized satellites, and helping to bridge the gap between theoretical embedded research and practical implementation in the space industry.
	While remaining flexible, and inducing only a minimal performance overhead, the presented multi-stage approach offers time-bounded real-time guarantees.
	
	The approach can be well complemented with several other reliability-improving measures which were integrated into the outlined reference MPSoC architecture.
	Preliminary benchmark results of an unoptimized implementation show a low performance overhead, suggesting a beyond factor-of-5 performance increase over state-of-the-art radiation-hardened processors for space use.
	Our approach allows the host platform to scale vertically (more powerful processor cores and more interfaces per tile) as well as horizontally (more tiles), with virtually any modern processor core.
	Thereby, we aim to increase acceptance for software-side FT approaches in the space industry, building trust in hybrid hardware-software architectures.
	Thus, our approach is the first integral, real-world solution to enable the fault-tolerant application with modern MPSoC designs for critical satellite control applications, thereby enabling the use of such SoCs in future high-priority space missions.
	\bibliographystyle{IEEEtran}
	\bibliography{ats2017}
	
\end{document}